\documentclass[12pt]{iopart}

\usepackage{iopams}
\usepackage{graphicx}
\usepackage{amsfonts}
\usepackage{amsbsy}
\usepackage{amssymb}
\usepackage{color}

\begin{document}

\title[Ratchet effect in an aging glass]{Ratchet effect in an aging glass}

\date{July 12, 2010}

\author{Giacomo Gradenigo}
\address{CNR-ISC and Dipartimento di Fisica, Universit\`a Sapienza - p.le A. Moro 2, 00185, Roma, Italy}

\author{Alessandro Sarracino}
\address{CNR-ISC and Dipartimento di Fisica, Universit\`a Sapienza - p.le A. Moro 2, 00185, Roma, Italy}

\author{Dario Villamaina}
\address{CNR-ISC and Dipartimento di Fisica, Universit\`a Sapienza - p.le A. Moro 2, 00185, Roma, Italy}

\author{Tom\'as S.\ Grigera} \address{Instituto de Investigaciones
  Fisicoqu{\'\i}micas
  Te{\'o}ricas y Aplicadas (INIFTA), and Departamento de F{\'\i}sica, Facultad de Ciencias Exactas,
  Universidad Nacional de La Plata, and CCT La Plata, Consejo
  Nacional de Investigaciones Cient{\'\i}ficas y T{\'e}cnicas,
  c.c. 16, suc. 4, 1900 La Plata, Argentina}

\author{Andrea Puglisi}
\address{CNR-ISC and Dipartimento di Fisica, Universit\`a Sapienza - p.le A. Moro 2, 00185, Roma, Italy}

\ead{ggradenigo@gmail.com, andrea.puglisi@roma1.infn.it}


\begin{abstract}
We study the dynamics of an asymmetric intruder in a glass-former
model. At equilibrium, the intruder diffuses with average zero
velocity. After an abrupt quench to $T$ deeply under the mode-coupling
temperature, a net average drift is observed, steady on a logarithmic
time-scale. The phenomenon is well reproduced in an asymmetric version
of the Sinai model. The subvelocity of the intruder grows with
$T_{eff}/T$, where $T_{eff}$ is defined by the
response-correlation ratio, corresponding to a general
behavior of thermal ratchets when in contact with two thermal
reservoirs.
\end{abstract}

\maketitle

In an irreversible environment, thermal fluctuations can be rectified
  in order to produce a directed current. After a few fundamental
  examples of historical and conceptual value~\cite{SMOL12,FEY1}, in
  the last twenty years a huge amount of devices and models ---usually
  known as Brownian ratchets or motors--- have been
  proposed~\cite{97SCIENCEastu,R02,HM09}. The purpose of these models
  is often practical, e.g. the extraction of energy from a highly
  fluctuating environment, such as a living
  cell~\cite{94NATURErous,10PNASDIL}. But Brownian ratchets are also
  valid probes for the non-equilibrium properties of the fluctuating
  medium, the value of the current being sensitive to the interplay of
  different time-scales as well as different temperatures at
  work~\cite{RBHH96,K07}.

The simultaneous breaking of space and time-reversal symmetries is a
necessary condition to observe directed motion.  A classical example
of this mechanism (see for instance \cite{97SCIENCEastu}), the
so-called flashing ratchet, consists of a particle undergoing standard
overdamped diffusion and also subject to a zero average
space-asymmetric (e.g. sawtooth) potential. If this potential is
switched on and off according to a random/periodic time sequence, a
current of particles (steady on average) can be observed. This current
is driven by the energy injected into the system when switching on the
potential.

Another way of breaking the time-reversal symmetry is obtained by
coupling the system with reservoirs at two different temperatures: an
asymmetric intruder in such a multi-temperature environment displays
an average drift, performing as a Brownian motor. In many-body systems
this has been done, for instance, in~\cite{BKM04,CPB07}.

The present work, inspired from the latter scenarios with a continuous
flow of energy in a non-thermalized medium, shows a study of the
ratchet phenomenon in the aging dynamics of fragile glass-formers. It
is well known that such systems, when quenched below their
Mode-Coupling temperature, display an out-of-equilibrium dynamics
customarily described within a two-temperature scenario
\cite{PRE97TEFF,BCKM98,00JPCMkob,ZBCK05}.  Fast modes are equilibrated
at the bath temperature while slow modes remain at a higher effective
temperature $T_{eff}$.  Here we show that the energy flowing from slow
to fast modes can be rectified to produce a directed motion.  The
properties of the observed current characterize the non-equilibrium
behavior of the glass. In particular a striking monotonic relation is
observed between the ratchet sub-velocity and $T_{eff}/T$. 
The experience with other kinetic models of
ratchets~\cite{BKM04,CPB07} teaches that - when in the presence of a
temperature unbalance - the heat flux also governs the ratchet
velocity.  These observations suggest the conjecture that one can obtain
the effective temperature by replacing the measure of linear response
and correlations with the simple measure of an average current. In
what follows we test the reliability of this procedure.

The first of the numerical experiments proposed here involves the 
3D soft-spheres model, which is a well known fragile glass-former
\cite{JPCM89SS,PRL97fdr,JCP99COLUZZI,PRL02GEO}.  The
thermodynamic properties of soft-spheres are controlled by a
single parameter $\Gamma=\rho T^{-1/4}$, which combines the
temperature $T$ and the density $\rho$ of the system, with $\rho = N/V
\sigma_{0}^{3}$ and $\sigma_{0}$ the radius of the effective
one-component fluid.  We study a binary mixture (50:50) with radii
ratio, $\sigma_{1}/\sigma_{0}=1.2$.  Particles interact via the soft
potential $U(r) = [(\sigma_{i}+\sigma_{j})/r]^{12}$ and the 
dynamics is evolved via a local Monte Carlo algorithm.  
Time is measured in Monte Carlo steps (one steps corresponds to 
$N$ attempted MC moves).   
The model has a dynamical crossover at a mode-coupling  temperature
$T_{MC}$ corresponding to the effective coupling $\Gamma_{MC}=1.45$
\cite{JPCM89SS}.  
The ratchet is formed by the asymmetric interaction of 
a single particle, the intruder, with all the other particles of the system.
Denoting with $x_i,$ $i>0$ the $x$ coordinate of the $i$-th particle,
and with $x_0$ the abscissa of the intruder, we choose

\begin{equation}
U({\bf r}_0,{\bf r}_i) = \left\{
\begin{array}{rl}
U(|{\bf r}_0-{\bf r}_i|) & \textrm{if $x_i < x_0$},\\
\varepsilon U(|{\bf r}_0-{\bf r}_i|) & \textrm{otherwise,}
\end{array} \right.
\label{UU}
\end{equation}
with $\varepsilon=0.02$.  The spatial symmetry along the $x$-axis is
therefore broken, fulfilling one of the requirement to get a ratchet
device. Let us stress that here, in analogy with the
model proposed in~\cite{BEB09}, the asymmetry is inherent to a
\emph{single} object, the intruder, which is embedded in a symmetric
environment, at variance with flashing and rocket ratchets~\cite{HM09}.
Moreover, by exploiting condition~(\ref{UU}), we are able to build an intruder with
an intrinsic asymmetry, even if it is a point-like particle.

We equilibrate configurations embedding an asymmetric particle at a
high temperature $T_{liq} \gg T_{MC}$. At $T_{liq}$ the system has
simple liquid behaviour with fast exponential relaxation.  The
dynamics of the asymmetric intruder is then studied both along
equilibrium trajectories and after quenches to different temperatures
$T<T_{MC}$. The average displacements of the intruder along different
axes is compared at and outside equilibrium, namely both $\langle
\Delta x_0(t) \rangle$ and $\langle \Delta y_0(t) \rangle$ are
studied, with $\Delta x_0(t) = x_0(t)-x_0(t_0)$. In
particular $t$ is the time elapsed since the instantaneous quench,
namely we take $t_0=0$.  The averages denoted
by $\langle \ldots \rangle$ are realized considering $5000$ initial
configurations with a single intruder, equilibrated at $T_{liq} =
4.42\, T_{MC}$, each followed by an independent thermal history.
Quenches are done to $T=0.67 \,T_{MC}$, $0.53\, T_{MC}$, $0.42\,
T_{MC}$, $0.31\, T_{MC}$.

\begin{figure}[ht!]
\centering
\includegraphics[width=0.75\textwidth,clip=true]{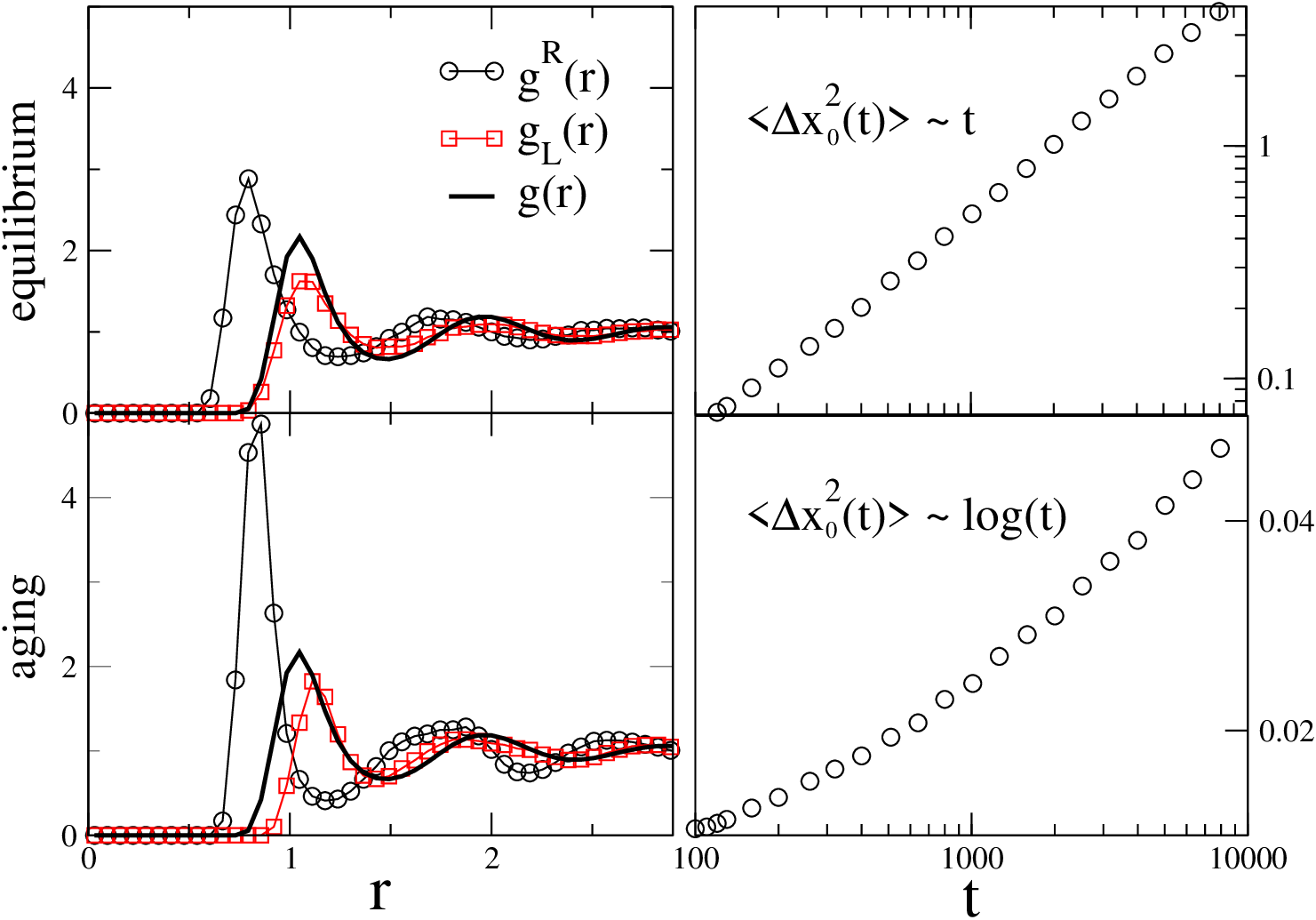}
\caption{Left panels: pair distribution function centered
    on the intruder, $g^{R}_{L}(r|\delta x \gtrless 0)$, of left (red squares),
    and right (black circles) neighbors, at equilibrium (top) and 
    at a fixed elapsed time after quench (bottom). 
    Black line, $g(r)$ for symmetric interactions.
    Right panels: $\langle \Delta x_0^{2}(t) \rangle$ at equilibrium  (top) and 
    after the quench} (bottom).
\label{fig:gr}
\vspace{-0.2cm}
\end{figure}

Let us consider first the effect of the asymmetric interaction on the
arrangement of particles around the intruder.  
We study the pair distribution function of neighbors to the right and
to the left of the intruder, 
$g^{R}_{L}(r|\delta x \gtrless 0) =
\sum_{j|\delta x_j \gtrless 0} \delta(r_{j0}-r)$, with $r_{j0}$ the distance 
between the intruder and the $j$-th particle, in equilibrium or
at a fixed elapsed time after the quench (see Fig.~\ref{fig:gr}, left panels).
Right neighbors stay closer to the intruder due to the reduced
repulsion.  This asymmetric clustering of neighbors is slightly
enhanced during aging, but at this level of analysis there is no
qualitative differences between the equilibrium and
off-equilibrium regimes.

\begin{figure}[ht!]
\centering
\includegraphics[width=0.75\textwidth,clip=true]{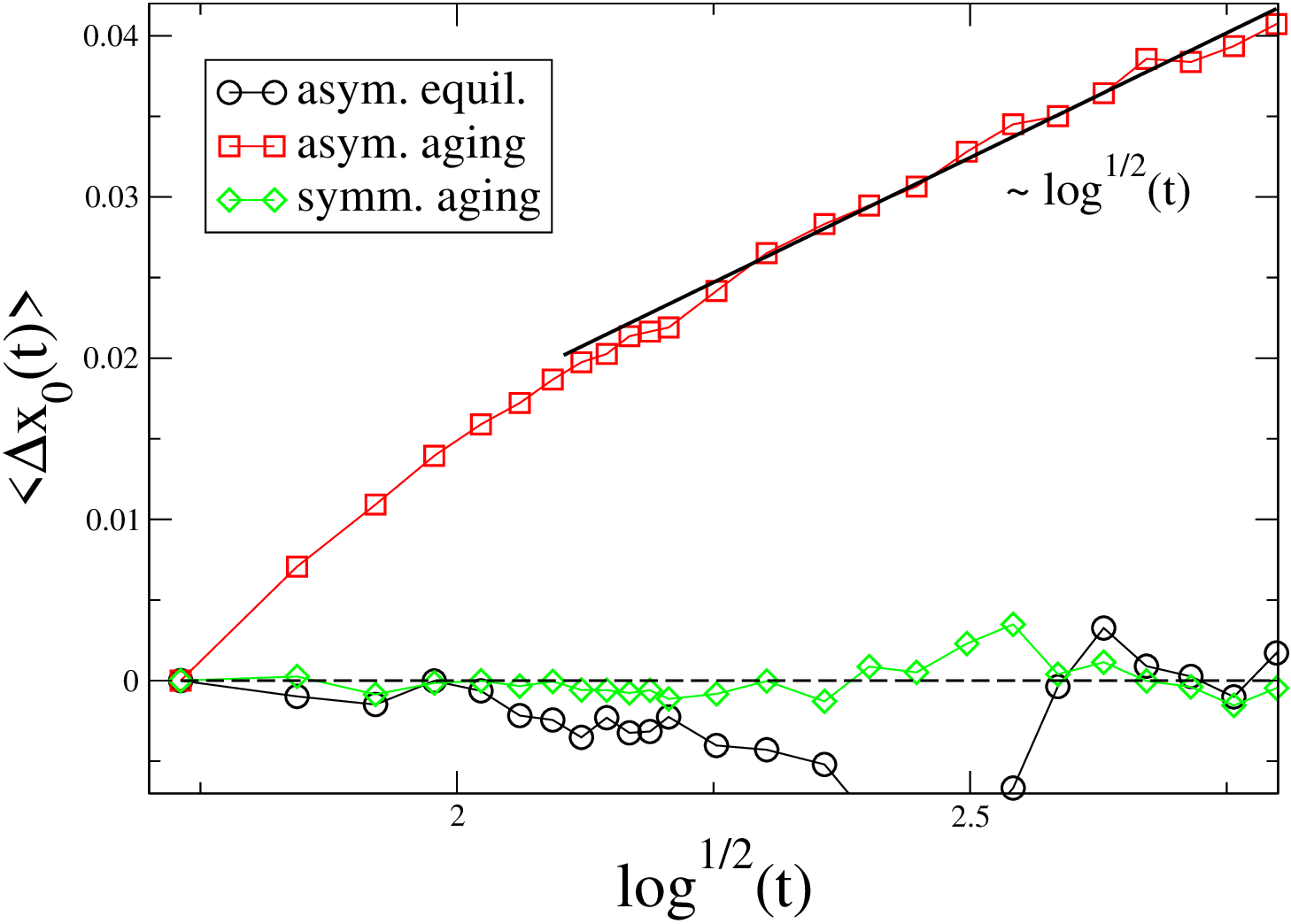}
\caption{Average intruder displacement:  
  $\langle \Delta x_0(t) \rangle$ for equilibrium trajectories (black circles), and 
  $\langle \Delta x_0(t)\rangle$ and $\langle \Delta y_0(t)\rangle$ after a quench far
    below $T_{MC}$ ($T = 0.42 T_{MC}$) (red squares and green diamonds respectively).
    The displacement $\langle \Delta x_0(t) \rangle$ is measured in units of the average inter-particle distance,
    as obtained from the position of the first peak in the pair distribution function $g(r)$ 
    (see also fig.~\ref{fig:gr}). 
}
\label{fig:drift}
\vspace{-0.2cm}
\end{figure}

Dynamical measurements, on the contrary, show important differences
between the two regimes. Let us consider the mean square
displacement around the average position at time $t$ of the intruder
particle (Fig.~\ref{fig:gr} right): it is linear with time at high
temperatures, while it is logarithmic after a quench, $\langle \Delta
x_0^2(t)\rangle-\langle\Delta x_0(t)\rangle^2 \sim \log t$. This
behavior, typical of activated dynamics in a rough potential
~\cite{92bouchWEB,PRL97fdr,PRL01APR}, is also observed for host
particles.  The behavior of the average \emph{displacement} of the
intruder, $\langle \Delta x_0(t) \rangle$, is more striking.  At
equilibrium (Fig.~\ref{fig:drift}, black circles) there is no net
displacement: this is because parity along $x$ axes is broken whereas
time reversal symmetry is preserved.  The same happens to $\langle
\Delta y_0(t) \rangle$ (green diamonds) during aging: in this case
only (macroscopic) time reversal symmetry is broken.  But parity and
time-reversal symmetry are both violated for $\langle \Delta x_0(t)
\rangle$ after the quench (red squares), and in this case a net
average drift is found (Fig.~\ref{fig:drift}), linear on a logarithmic
timescale, $\langle \Delta x_0(t) \rangle \sim \tau$ with $\tau
=\log^{1/2}t$.  The logarithmic timescale again points to a
non-equilibrium phenomenon ruled by activated events.  
Moreover, the simple scaling relation $\langle \Delta x_0(t)\rangle
\sim \sqrt{\langle \Delta x^2_0(t)\rangle-\langle \Delta
x_0(t)\rangle^2}$ between the displacement along the $x$ axis and the
m.s.d. around the average position is observed, in analogy with the Sinai
model discussed below.

\begin{figure}[ht!]
\centering
\includegraphics[width=0.45\textwidth,clip=true]{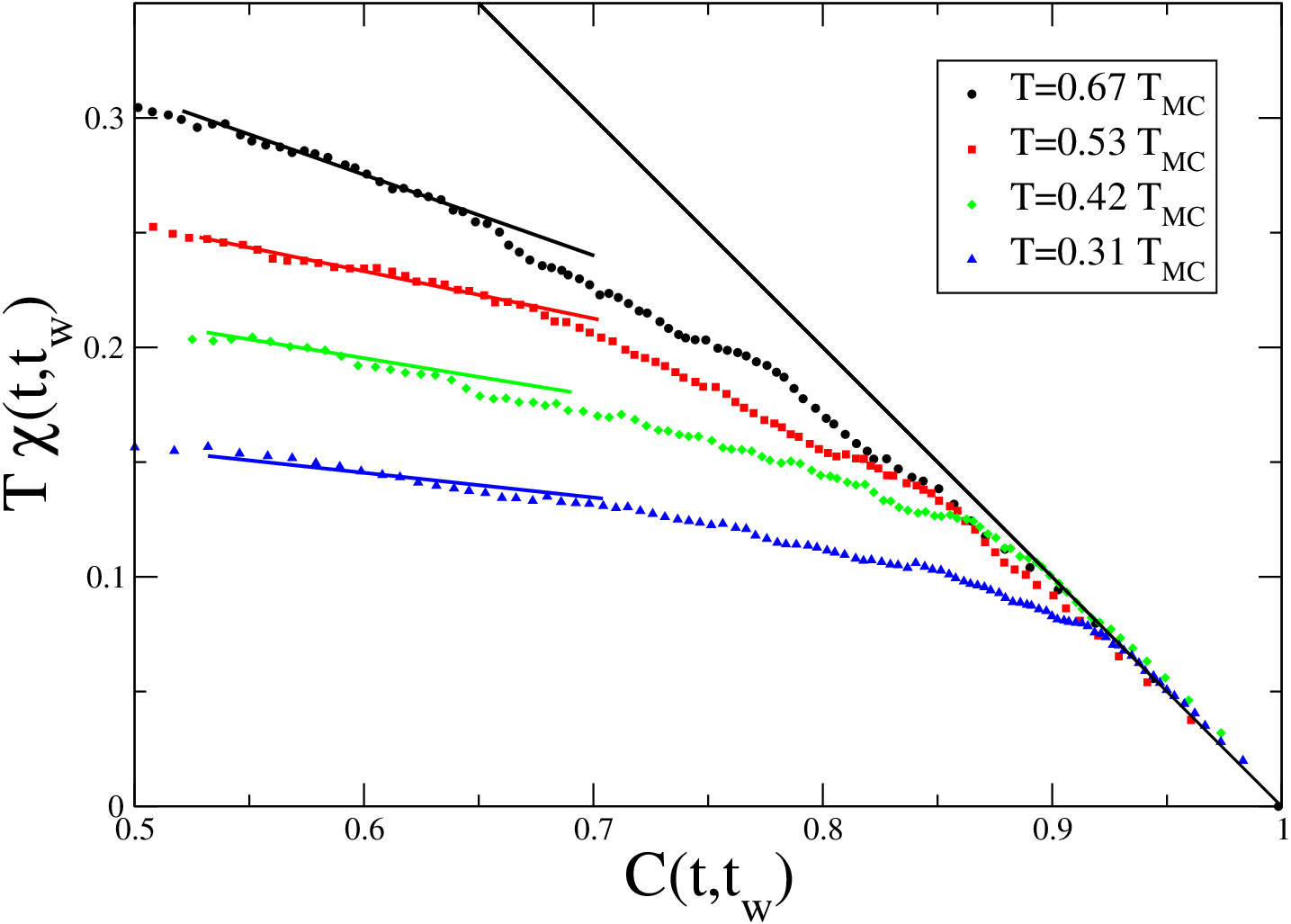}
\includegraphics[width=0.45\textwidth,clip=true]{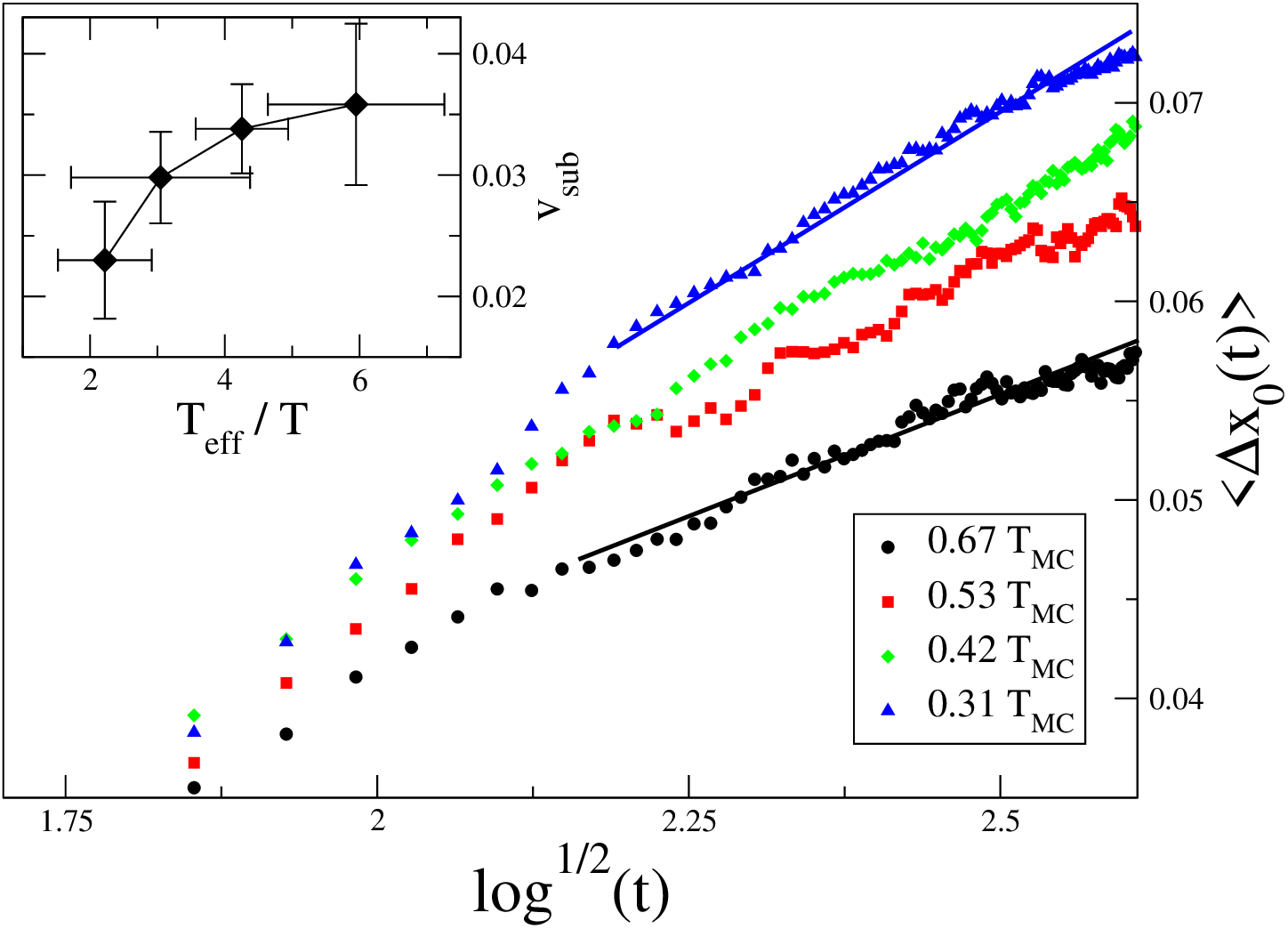}
\caption{Left panel: parametric plot $T\chi(t,t_w)$ \emph{vs}
$C(t,t_w)$.  Right panel: asymmetric intruder drift for different
quench temperatures; linear fits yield sublinear velocities.  Inset:
$v_{sub}$ \emph{vs} $T_{eff}/T$.  In both panels colors of
data correspond to different effective tempeartures:
$T_{eff}=1.497~T_{MC}$ (black), $1.613~T_{MC}$ (red), $1.792~T_{MC}$
(green) and $1.819~T_{MC}$ (blue).}
\label{fig:tempdrift}
\end{figure}

Our first exploration of ratcheting effects in glassy models leaves
open one important question, namely the dependence of the drift on the
parameters (e.g.\ temperature) of the system. In order
to sketch a preliminary answer, we need to extract from the curve
$\langle \Delta x_0(t) \rangle$ a synthetic observable. 
The finding of a
logarithmic timescale $\langle \Delta x_0(t) \rangle \sim \tau$ with
$\tau \sim \log^{1/2}t$ 
suggests to define an average \emph{sub-velocity} as~\cite{G10}:
\begin{equation}
v_{sub}(t,t_w)=\frac{\langle \Delta x_0(t) \rangle - \langle \Delta x_0(t_w) \rangle}{\delta \tau} =
\frac{\langle x_0(t) \rangle - \langle x_0(t_w) \rangle}{\delta \tau},
\end{equation}
with $\delta \tau=\log^{1/2}t-\log^{1/2}t_w$. This average subvelocity
depends in general on both the running time $t$ and the waiting time $t_w$ elapsed
since the quench. More precisely, considering fig.~\ref{fig:drift}, by fixing $t_w$ and $t$ 
we choose the time lag where the slope of the curve $\langle \Delta x_0(t) \rangle$ is measured. 
Clearly, the \emph{instantaneous} sub-velocity only depends on $t_w$
and corresponds to a slowly decaying
velocity $v(t_w)=d \langle \Delta x_0(t_w) \rangle/d t_w \sim 1/t_w$ (with
logarithmic corrections).  
For large enough waiting times we can define the ``order parameter'' $v_{sub}$, 
namely we find $v_{sub}(t_w) \sim \textrm{const}$, and then we
probe its dependence upon the external parameters. Here we focus on
the quench temperature $T$, drawing a connection between the drift of
the glassy ratchet, which is a pure non-equilibrium effect, and the
more customary equilibrium-like descriptions of aging media in terms
of effective temperatures. In the right panel of
Fig.~\ref{fig:tempdrift} we show the drift of the asymmetric intruder
for four quench temperatures: $T=0.67 T_{MC}$, $0.53 T_{MC}$, $0.42
T_{MC}$, $0.31 T_{MC}$.  We observe that when {\em decreasing} the
quench temperature the drift {\em grows} in intensity, i.e. the
sub-velocity increases. This somehow is a stronger evidence that the
ratchet drift cannot be described as an equilibrium-like effect,
e.g. trying to connect average kinetic or potential energy with a
mobility: one would expect that at lower temperatures everything is
slowed down, whereas we find that the ratchet drift is enhanced.

Typical examples of ratchets in (ideally) statistically stationary
configurations are obtained by coupling the system with two or more
reservoirs.  It is the existence of different temperatures within the
same system which allows the production of work without violations of
the second principle of thermodynamics.  But what is the {\em second
temperature} in a glassy system?  According to the well-established
description of the aging regime of glasses \cite{BCKM98}, it is the
effective temperature defined as the violation factor of the
fluctuation-dissipation theorem (FDT). The quench of a fragile glass,
for instance our the soft spheres model, below its mode-coupling
temperature produces aging and violations of FDT.  The factor
$X(t,t_w)<1$ which allows one to write a generalized FDT,
$T\chi(t,t_w)= X(t,t_w)[C(t,t)-C(t,t_w)]$, with $\chi(t,t_w)$ the
integrated response and $C(t,t_w)$ the correlation, yields the
definition of the effective temperature $T_{eff}(t,t_w)=T/X(t,t_w)$.
The last is usually higher than the bath temperature $T_{eff}(t,t_w) >
T$ and is understood as the temperature of slow, still not
equilibrated, modes.  Clearly, the ratio $T_{eff}/T$ may
be regarded as the parameter which tunes non equilibrium effects and
we study here how the glassy ratchet drift depends on it.  We obtain
$T_{eff}$ from the parametric plot of $C(t,t_w)$, taken as the
self-intermediate scattering function, versus the integrated response
$T \chi(t,t_w)$, measured according to the field-free method
of~\cite{PRL07BERTH}.  The parametric plot $T \chi(t,t_w)$ \emph{vs}
$C(t,t_w)$ is shown in the left panel of fig.~\ref{fig:tempdrift} for different temperatures. 

At large times $v_{sub}$ exhibits finite-size effects, namely the
drift saturates at a time that can be increased by increasing the size
of the simulation box. This is why we compute $v_{sub}$ by fitting
data only in a relative early time window, namely $10<t_w<10^{3}$.
Accordingly, the measure of $X(t,t_w)$ is obtained from the parametric
plot $T \chi(t,t_w)$ \emph{vs} $C(t,t_w)$ with $t=10^{3}$ and $t_w$
ranging from $10$ to $10^{3}$, namely the early aging regime is
considered. The inset of the right panel of fig.~\ref{fig:tempdrift} shows the behavior
of $v_{sub}$ \emph{vs} $T_{eff}/T$, revealing that the subvelocity
increases when $T_{eff}/T$ is increased. Namely the intensity of the
ratchet effect, traced in the measure of $v_{sub}$, grows as the
distance from equilibrium is increased. 

Let us remark here that the mechanism governing our ratchet differs
from that of a flashing ratchet~\cite{97SCIENCEastu}. Irreversibility
is achieved by choosing an \emph{initial condition} which is
out-of-equilibrium with respect to the bath temperature $T_{liq} \neq
T$. The extreme slowness of an aging glass prevents the system from
thermalizing, so that energy continuously flows from fast to slow
modes, supplying power to the Brownian ratchet.  At variance with a
flashing ratchet, here there is no external time-dependent modulation
of a potential. The movement of surrounding molecules produces
fluctuations of the intruder potential, but these obey a globally
conservative dynamics and cannot produce energy.

Indeed, simpler models can be conceived, where the intruder moves in a
  time-{\em independent} potential: as a enlightening example, in the
  following we discuss the Sinai model, where it is made clear how
  irreversibility comes \emph{solely} from the choice of initial
  conditions.  Sinai model is one of the simplest describing the
  diffusion of a single particle through a random correlated
  potential~\cite{PhR90BOUCH}.  Its long-time dynamics is ruled by
  activated events and is characterized by a logarithmic time-scale:
  for this reasons it appears to be a well fitted candidate to
  reproduce the previous experiment in a more controlled setup.  In
  the original Sinai model the random potential is built from a
  random-walk of the force on a $1d$ lattice. The force $F_i$ at each
  lattice site $i$ is an independent identically distributed random
  variable extracted from a zero mean \emph{symmetric} distribution
  $p(F)$. The potential on a lattice site $n$ is given by
  $U(n)=\sum_{i=1}^{n} F_i$.  Because {\bf $\int dF p(F) F = 0 $} the
  average force experienced by a particle is zero.  The potential
  excursion between two sites grows like $\langle |U(i)-U(j)| \rangle
  \sim |i-j|^{1/2}$.  The above relation, together with the expression
  of the typical time needed to jump a barrier, $\sim \exp(\beta
  \Delta U)$, yields the long-time scaling of the mean square
  displacement of: $\langle \Delta l^2(t)\rangle \sim \log^4(t)$, with
  $l(t)$ denoting the site occupied by the particle at time $t$ and
  $\Delta l(t)=l(t)-l(0)$. A logarithmic time scale for the growth of
  domain size is quite ubiquitous in the low temperature regime of
  glassy systems, where activated processes dominate~\cite{CCY10}.
  The time-reversal symmetry breaking is inherent in the Sinai model:
  by averaging over initial positions extracted from a flat
  distribution one reproduces an initial infinite temperature
  $T_{liq}=\infty$, while the quench to a glassy phase is reproduced
  evolving the system at temperature $T \ll \sqrt{L}$ where $L$ is the
  linear size of the system.  We propose a {\em spatially asymmetric}
  version of the Sinai model, where the symmetric force distribution
  $p(F)$ is replaced with an asymmetric one $\tilde{p}(F)$, in analogy
  with the asymmetric potential of the glassy ratchet, in order to
  find a non-equilibrium drift.  Namely the distribution
  $\tilde{p}(F)$ is such that $\tilde{p}(F)\neq \tilde{p}(-F)$, but
  still $\int dF \tilde{p}(F) F = 0 $.  In particular, the random
  force is obtained according to the following procedure: at each site
  the sign of a random variable $f_i$ is chosen with probability $1/2$
  and its modulus, in agreement with the sign, from either
  $e^{-|f_i|/\lambda_{-}}$ or $e^{-|f_i|/\lambda_{+}}$, with
  $\lambda_+ > \lambda_-$. In the present simulation $\lambda_+=1$ and
  $\lambda_-=0.2$ have been used.  The potential is then built as
  $U(n)=\sum_{i=1}^{n} F_i$, with $F_i=f_i-(\lambda_+-\lambda_-)/2$.
  This amounts to a shift of the whole distribution such that $\langle
  F \rangle=0$.

\begin{figure}[ht!]
\centering
\includegraphics[width=0.75\textwidth,clip=true]{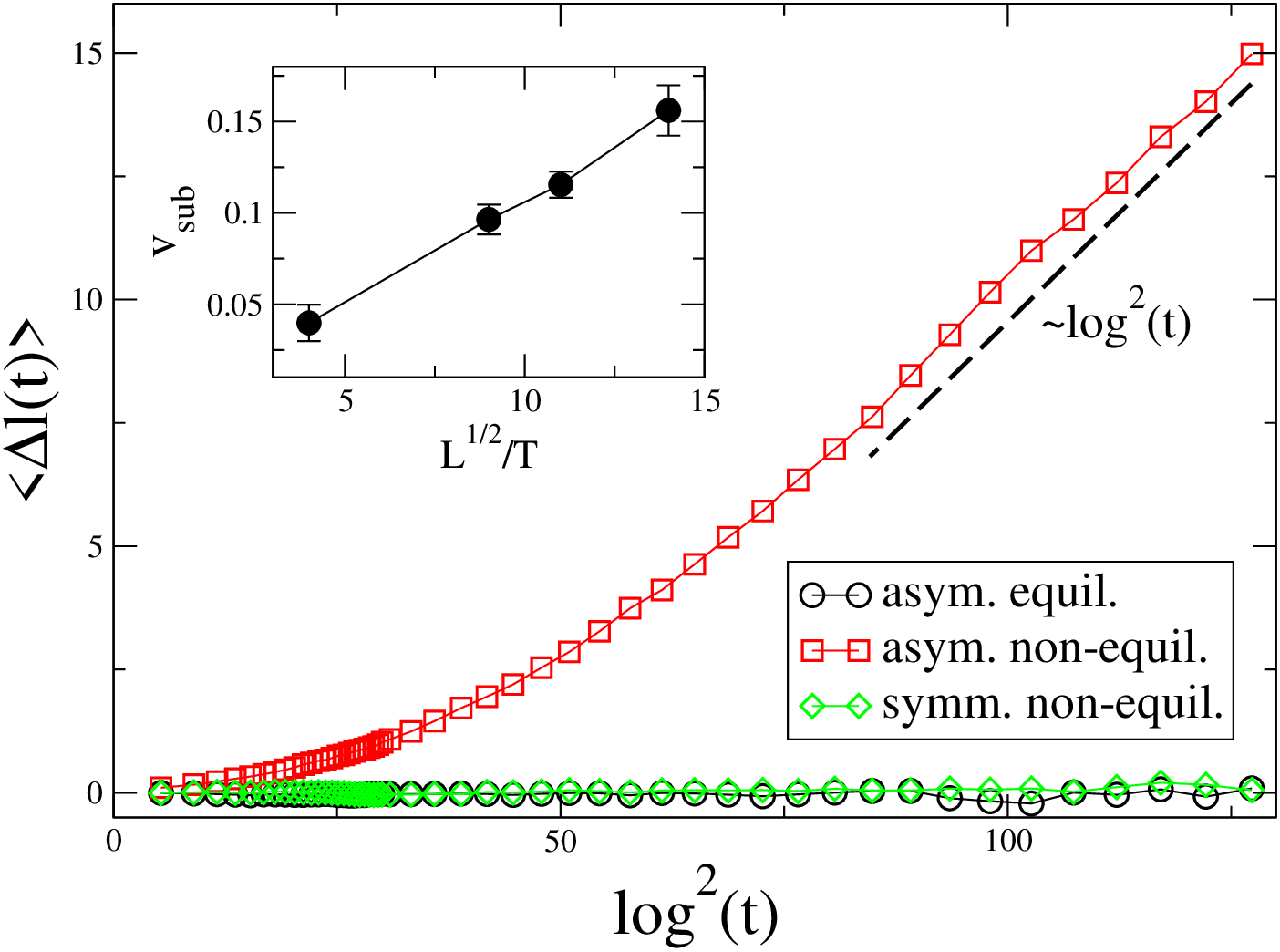}
\caption{Main: Average displacement in Sinai
model. Black circles, diffusion in asymmetric potential at
equilibrium; green diamonds, diffusion during aging in symmetric
potential (quench from $T=\infty$); red squares, diffusion during aging
in asymmetric potential. Inset: the sublinear drift velocity $v_{sub}$
grows with $\sqrt{L}/T$.}
\label{fig:sinai}
\vspace{-0.2cm}
\end{figure}

The results of the simulations of this model are shown in
Fig.~\ref{fig:sinai}: they clearly show a behavior in striking
similarity with those of Fig.~\ref{fig:drift} for the glass-former
model: a drift is observed only when both time reversal and spatial
symmetry are broken.  Indeed, as demonstrated by the black and red
solid curves, it is sufficient that one of the two symmetries is
restored to have a zero drift. In particular, to obtain an
``equilibrium'' dynamics in this model it is sufficient to distribute
all initial positions of the simulation according to a distribution
$\sim\exp[-\beta U(i)]$: in this case, even with the asymmetric
distribution of forces described above, the average drift is zero.
The large time behavior of the drift is compatible with a squared
logarithm, $\langle \Delta l(t) \rangle \sim \log^2(t) \sim
\sqrt{\langle \Delta l^2(t) \rangle}$, in agreement with what
observed previously for the glassy ratchet.  

The out-of-equilibrium dynamics of the Sinai model closely reproduces
these observations: in the inset of Fig.~\ref{fig:sinai} shows the
sub-velocity as a function of $\sqrt{L}/T$, with $T \ll L^{1/2}$ the
quench temperature and $L$ the size of the linear chain, which fixes
the most relevant energy scale of the model, i.e.\ the maximum depth
of potential minima.  The inset of Fig.~\ref{fig:sinai} shows that
$v_{sub}$ grows monotonously with $\sqrt{L}/T$: when the last quantity
is increased particles condensate on the bottom of the deepest
valleys. The monotonic increase of $v_{sub}$ with $\sqrt{L}/T$ signals
that also for the Sinai model the larger the distance from equilibrium
the larger the velocity of the drift.

In conclusion, through numerical simulations in different models and
different choices of the quench temperature, always chosen in the deep
slowly relaxing regime, we have given evidences of the existence of a
``glassy ratchet'' phenomenon. The drift velocity slowly decays in
time and can be appreciably different from zero for at least three
orders of magnitude in time. The overall intensity of the drift,
measured in terms of a ``sub-velocity'', is monotonically increasing
with the distance from equilibrium, i.e.\ with the difference between
the quench and effective temperatures. This observation supports the
idea of regarding the ratchet drift as a ``non-equilibrium
thermometer'': it can be used as a device capable to say how far is a
system from equilibrium.  Nevertheless, for such a thermometer to be
effective, a more accurate calibration procedure should be carried
on. Namely, one should verify that $v_{sub}$ is a function of
$T_{eff}$ and $T$ only with a small number of parameters. In this case
the calibration of our thermometer would require a small
independent measures of $T_{eff}$ to fix those parameters.

As an experimental realization of our glassy ratchet, one
should consider a particle with anisotropic interaction with the
surrounding ones with a small magnetic dipole placed on it, orthogonal
to the asymmetry axis.  Then, switching on a constant magnetic field
the orientation of the two faces of the ``Janus''~\cite{MRL05} particle is
preserved with respect to a fixed reference frame, so that the spatial
symmetry of the interaction is broken.  Recent theoretical and
experimental advances in the study of functionalized or ``patchy''
particles~\cite{BLTZS06,GS07} promise an experimental
verification of our hypothesis in the near future.

\ack

We thank A.Baldassarri and A.Vulpiani 
for many useful discussions. The work of GG, AS, DV and AP is supported by the
``Granular-Chaos'' project, funded by Italian MIUR under the
grant number RBID08Z9JE. TSG was partially supported by
ANPCyT (Argentina).


\section*{References}
\bibliographystyle{unsrt}
\bibliography{paper.bib}

\end{document}